# Dynamics of interacting skyrmions in magnetic nano-track

W. Al Saidi,[1] R. Sbiaa,[1*] S. Bhatti,[2] S. N. Piramanayagam,[2] S. Al Risi,[1] and O. Al Bahri[1]

[1]Sultan Qaboos University, College of Science, Department of Physics, P.O. Box 36, PC 123, Muscat, Oman

[2]School of Physical and Mathematical Sciences, Nanyang Technological University, 21 4 Nanyang Link, Singapore, 637371

* Corresponding author

**Abstract**

Controlling multiple skyrmions in nanowires is important for their implementation in racetrack memory or neuromorphic computing. Here, we report on the dynamical behavior of two interacting skyrmions in confined devices with a comparison to a single skyrmion case. Although the two skyrmions shrink near the edges and follow a helical path, their behavior is different. Because the leading skyrmion is between the edge and the trailing one, its size is reduced further and collapses at a lower current density compared to the single skyrmion case. For higher current density, both skyrmions are annihilated with a core-collapse mechanism for the leading one followed by a bubble-collapse mechanism for the trailing one.

## I. Introduction

Magnetic skyrmions are chiral spin textures, which can exist in nanoscale size. Skyrmions can be displaced by spin-transfer torque (STT) with a current much smaller than that required for the domain walls [1–5]. Since their experimental observation in bulk materials [6,7] and multilayer thin films [8,9], skyrmions have been intensively investigated due to their motion at low energy, reasonably good velocity (up to 100 m/s), and ability to circumvent material defects [4,9–11]. Because of these characteristics, a dense skyrmion state has been considered propitious for multistate memory and neuromorphic computing devices [12–16]. Such applications rely on moving multiple skyrmions in a track. However, the skyrmion motion in a confined geometry involves different complex interactions, which causes a skyrmion to deviate from the path, change in size, reshape, annihilate, and repel from the track-edges [17–24]. In addition, the dynamics of multiple skyrmions become more complicated due to the skyrmion-skyrmion interactions. Therefore, the collective behavior of skyrmions within a nano-track needs to be understood thoroughly prior to realizing any skyrmion-based application.

In this study, we report the current-driven dynamical behaviors of two interacting skyrmions in a confined geometry and evaluate them against the dynamics of a single skyrmion. We found that the dynamics of skyrmions due to the repulsive interaction force could generate several effects, including compression, swirling motion, and annihilation. Additionally, we realized that the skyrmions experience two different types of annihilations: core-collapse and bubble-collapse. Although both annihilations occurred at a timescale of pico-seconds, the former annihilation mechanism was faster than the latter. Furthermore, we report an interesting interplay between two skyrmions during the current-driven motion, which results in their different behaviors.

Our results facilitate the understanding of the interaction between skyrmions driven by STT effect in a confined geometry. Further, this study highlights the difference between the behaviors of two skyrmions and can provide guidelines for the control of skyrmion flows in racetrack memory and neuromorphic computing devices.

## II. Results

**II. 1. Single skyrmion dynamics.** We used a single-layer effective model to perform the simulations (see the Simulations Section). The dimensions (L × W × H) of the simulated nano-track were $300 \times 70 \times 2$ nm$^3$. We started with nucleating a Néel skyrmion (hedgehog) at the center of the device with the topological number of N = -1. Intrinsic material parameters (mentioned in the simulation section) determined the skyrmion radius to be about 13.5 nm as previously reported [25]. We calculated the radius of the skyrmion by measuring the distance from the core to the in-plane component of its magnetization.

An electric current with a density $J$ and a duration of 20 ns was applied from left to right to move the skyrmion. For small values of $J$ ($< 2 \times 10^{11}$ A/m$^2$ in our case), the skyrmion will not be able to reach the right edge and bounce back after removing the current. For $J$ between $2 \times 10^{11}$ A/m$^2$ and $10 \times 10^{11}$ A/m$^2$, the skyrmion performed a swirling motion (Figs. 1a-b). We noticed a decrease in the radius of skyrmion with the increasing $J$. Finally, the large values of $J$ ($> 10 \times 10^{11}$ A/m$^2$), resulted in the annihilation of skyrmion.

Fig. 1b shows the skyrmion trajectory near the right edge of the nanowire for $J = 5 \times 10^{11}$ A/m$^2$ and $8 \times 10^{11}$ A/m$^2$. The position and trajectory of the skyrmion was measured from its core. We realized a reduction in the skyrmion size $R$ to 8.4 nm (represented by the color code in Fig. 1b). Furthermore, we realized that the radius $R$ changes depending on the skyrmion distance to both right and top edges and not only the right one as reported. We defined $d_{x,m}$ and $d_{y,m}$ as the minimum distances of the skyrmion to the right and top edges, respectively. The simulation showed that the minimum radius is obtained closer to the corner of the nanowire and not nearer to any of the two edges. Moreover, we noticed that the skyrmion swirling becomes larger as the applied current increases. From Fig. 1c it can be seen that the minimum distance between the skyrmion position and the top corner $r_m$ is inversely proportional to $J$ while the distances of the skyrmion to the right edge $d_{x,m}$ and to the felt edge $d_{y,m}$ show exponential decay behaviours. Moreover, we noticed a repulsion of skyrmion from the right edge due to the presence of the tilted magnetic vectors at the edge, also known as the edge states [26–29]. The swirl motion and reduced size of the skyrmion at the edge is a repercussion of competition between multiple forces acting on the skyrmion. As the skyrmion is under the STT effect in the right direction and the resultant force from the two edges (Fig. 2), its size is reduced when it is strongly driven to the right (high $J$ values). The interaction force from the two edges is defined as:

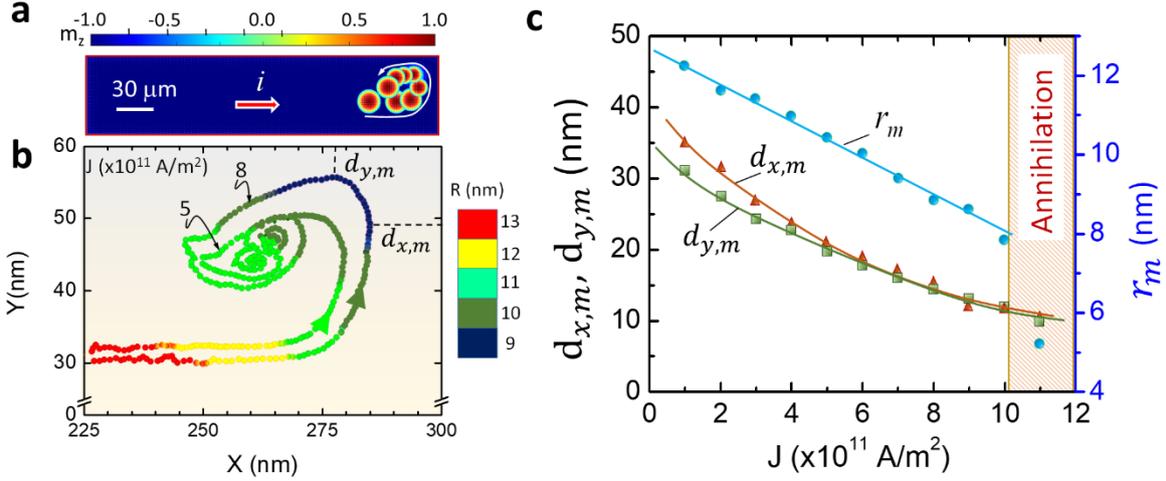

**Figure 1 | Single skyrmion dynamics.** (a) The motion of the skyrmion near the edge of the nanotrack at $J = 5 \times 10^{11}$ A/m² where a change of the size can be observed. It is clearly seen that the minimum size of the skyrmion is obtained near the corner due to the addition of the forces from the two edges. The nanowire dimensions are $L = 300$ nm and $W = 70$ nm. (b) Skyrmion motion trajectory experiencing a spiral motion for $J = 5 \times 10^{11}$ and $8 \times 10^{11}$ A/m². The color code indicates the radius of the skyrmion and (c) The minimum positions to the right edge $d_{x,m}$, top edge $d_{y,m}$ and the corner $r_m$ versus the current density.

$$\vec{F} = \vec{F}_R + \vec{F}_T \tag{3}$$

$$\vec{F} = -f_o e^{-\frac{d_x}{d_o}} \hat{\imath} + -f_o e^{-\frac{d_y}{d_o}} \hat{\jmath} \tag{4}$$

The first and second terms on the right of Eq. (4) are the repulsive forces from the right and top edges, respectively. The parameters $d_x$ and $d_y$ are the positions of the skyrmion from the right and the top edges, respectively. The unit vectors $\hat{\imath}$ and $\hat{\jmath}$ are along the *x*-axis and *y*-axis, respectively. The parameters $f_o$ and $d_o$ are materials dependent as described by Navau *et al.* [19]. The response of the skyrmion to repulsive forces was evidenced by the reduced size and the spiral motion to avoid annihilation. However, for a large driving force from the STT effect, the skyrmion became too small ($R < 5$ nm) and annihilated as shown in the dashed region of Fig. 1b. By knowing the maximal repulsive force from Eq. (2), the minimum radius $R_m$ can be obtained. If the driving force from STT is too large ($J > 10 \times 10^{11}$ A/m²), skyrmion overcomes the repulsive force from the edge resulting in its annihilation, since it loses its topological protection at the edge [30]. The critical current, at which skyrmions annihilate depends on the material parameters and the device geometry [9,21]. For smaller track width, the calculation showed that the resulting repulsion force is higher (higher compression of skyrmion by the edge) which will keep the skyrmion in the device and thus higher current is needed to annihilate it. The trajectory of the spiral motion is adjustable by external parameters such as the current, or internal parameters where the force can be strengthened or weakened. The skyrmion swirling motion direction is independent of the material parameters and it was found that the skyrmion deflection to the top or bottom depends on the sign of the skyrmion charge Q; i.e. for Q = -1, the direction is counterclockwise while for Q= +1 the swirling is clockwise direction. As the skyrmion moved far from the right edge, it expanded as a response to a reduction of $\vec{F}_R$. The change of $R_m$ versus $r$ (distance to the corner) is plotted in Fig. 2.a. A linear dependence of $R_m$ with the skyrmion distance from the corner is clearly observed for $r <$ 35 nm then a smaller change of $R_m$ can be noticed for $r >$ 35 nm indicating that the repulsion forces are weakened and the skyrmion expands to its initial size. As will be discussed later, the collapse of the skyrmion occurred at a position closer to the edge rather than the right edge. Fig. 2b is a plot of the magnitude of net repulsive force $\vec{F}$ as a function of the position r shown in inset of Fig. 2a. The force $F$ follows an exponential decay function ($F \propto e^{-r/r_o}$) where the parameter $r_o$ is 10.45 nm for $J = 8 \times 10^{11}$ A/m$^2$.

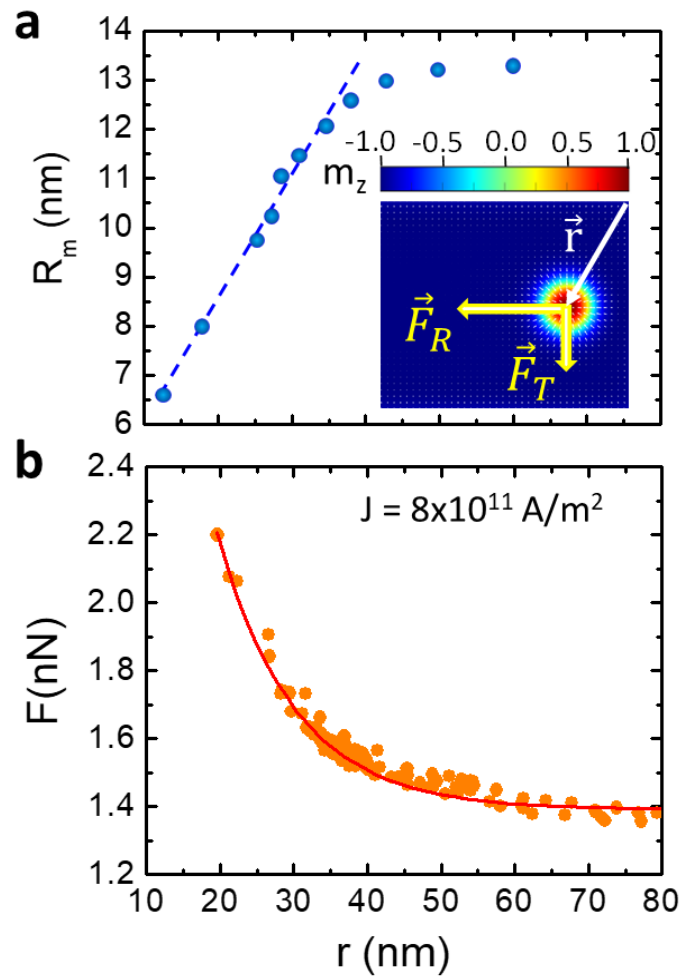

**Figure 2 | Dependence of the two repulsive forces on skyrmion size.** (a-c) Snapshots of the skyrmion showing the size and position at $J = 8 \times 10^{11}$ A/m² and for different calculation times. As the skyrmion is displaced under the STT effect, the magnitudes of the two repulsive forces are changed with a maximum strength near the corner. (b) The dependence of minimum radius $R_m$ of the skyrmion on its distance to the corner of the nanowire. The calculated net repulsive force is plotted as a function of the distance $r$ from the corner of the nanowire. It shows a good agreement with the exponential decay function. The color code indicates the z-component of the magnetization within the nanowire.

In addition to the change in the size of the skyrmion, its velocity $v$ shows an oscillatory behavior because of the repulsive force from the right and top edges and the driven force from the current to the right direction (Fig. 3.a). For $J = 8 \times 10^{11}$ A/m$^2$, a maximum velocity of about 50 m/s near the corner of the device with a minimum radius was observed when the skyrmion is too close to the corner before bouncing back in a helical path (state A). Similar results were obtained for $J = 5 \times 10^{11}$ A/m$^2$, with a maximum $v \sim$ 20m/s and larger skyrmion size at the corner (state B). The differences in $v$ and $R_m$ are due to lower driving current and thus the repulsive force are dominant. This is more clear for $J = 1 \times 10^{11}$ A/m$^2$, with a maximum $v$ of only 5 m/s and no change in the skyrmion size. The skyrmion acceleration is mainly due to the driven current, which is in this case very small compared to the net repulsive force from the edges. Figs. 3(b-d) are images of the skyrmion within the nanowire showing the states A, B and C indicated in Fig. 3.a. In state A under $J = 8 \times 10^{11}$ A/m$^2$, the skyrmion could be much closer to the corner where the size could be much smaller than $J = 5 \times 10^{11}$ A/m$^2$ (state B). This moderate value of $J$ was competing with repulsive force from the two edges and yet the skyrmion could see its size shrinking but less than in the first case. For a small $J$ of $1 \times 10^{11}$ A/m$^2$, the driving force from STT is not enough to displace the skyrmion closer to the edge and thus the size remains large (state C).

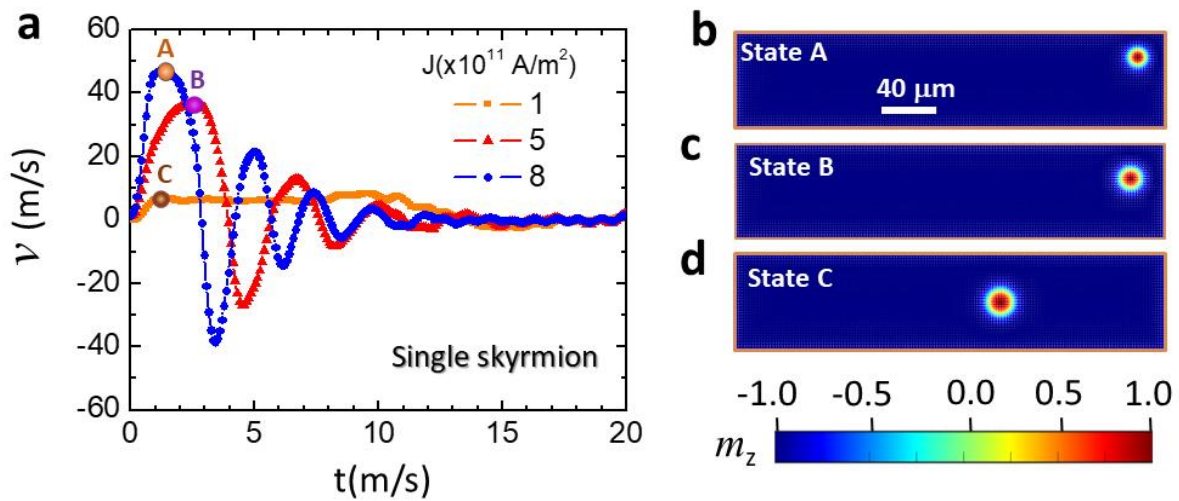

**Figure 3 | Velocity of a single skyrmion.** (a) The oscillatory behavior of skyrmion driven by STT effect under $J = 5 \times 10^{11}$ and $8 \times 10^{11}$ A/m$^2$. The maximum velocity is reached near the corner of the device with a strong reduction in size. For $J = 1 \times 10^{11}$ A/m$^2$, the skyrmion is moving with a low and constant velocity until stabilized far from the edges. (b-d) snapshots of the states A, B and C shown in Fig. 3a.

## II. 2. Two interacting skyrmions dynamics.

In the second part of this study, we introduced a second skyrmion (skyrmion-B) at a distance of 68 nm from the first skyrmion (skyrmion-A). The skyrmions were driven to the right of the nano-track using STT effect. Simultaneously, both skyrmions shrank when the distance between them decreases. A larger driving current will result in a stronger compression and thus, the annihilation of skyrmions [22,31]. Due to the skyrmion-skyrmion repulsion, the spacing between the skyrmions as well as their sizes changes with time. We calculated the spacing between two skyrmions by measuring the distance between their centers. We noticed a repulsion between the two skyrmions at short distances (smaller than 68 nm).
In addition, we noticed a reduction in the size of skyrmions as the separation $S$ was reduced (Fig. 4a). The reduction in size is explainable due to the multiple forces (STT, repulsive forces from neighboring skyrmion and edges) acting in opposite directions, which squeezes the skyrmions. Noticeably, the skyrmion-B reduces more in size than skyrmion-A, due to the repulsive forces from the right edge, which was missing for the skyrmion-A. Furthermore, we noticed a reduction in the minimum separation with the increase in the current densities as can be seen in Fig. 4b. This figure also shows the trend of skyrmions sizes as a function of current densities. For $J > 8 \times 10^{11}$ A/m$^2$, skyrmion-B, which is under two opposite repulsive forces, collapsed at the edge as indicated by zone 1 in Fig. 4b and supplementary Video S$_1$. The annihilation occurred due to large STT with a combination of the repulsive force from skyrmion-A overcoming the repulsive force from the two edges. The annihilation of skyrmion-B resulted in an expansion of skyrmion-A due to the vanishing of the repulsion force from skyrmion-B. Whereas increasing the current density further leads to annihilation of skyrmion-B as shown in zone 2 of Fig. 4b and supplementary Video S$_1$. Next, we tried to comprehensively understand the effect of skyrmion-skyrmion interaction. We plotted the path of two interacting skyrmions and compared it to the case of a single skyrmion. The comparison was done for an applied current density $J = 8 \times 10^{11}$ A/m$^2$ (shown in Fig. 5a). A single skyrmion (skyrmion-A) follows a helical path (red circle dots), as discussed previously. When skyrmion-A is not interacting with skyrmion-B, it reached a distance $d_{x,m} = 15.2$ nm closer to the right edge due to the repulsive forces from the two edges and the driving force from the STT. However, when it is interacting with skyrmion-B, its path is pushed further from the edge (cyan square symbols) with still a helical path. The change $\Delta x$ in the *x*-direction which is only due to skyrmion-skyrmion interaction reached a value of 34 nm. This mutual interaction affected also the path of skyrmion-B and it is manifested by the compression of its helical path and a slight shift toward the right side, opposite to the shift of skyrmion-A (blue symbols). This compression is the result of the resultant forces from the two edges and the interaction with skyrmion-A. Moreover, we calculated $\Delta x$ for different values of $J$ and it was found to be reduced monotonically with an increase in $J$ (Fig. 5.b). Skyrmion-A is driven by STT and also repealed from skyrmion-B. Still the interaction with skyrmion-A is dominant but as $J$ is becoming larger, STT competes with this interaction and that is the reason why skyrmion-B is annihilated for $J > 8 \times 10^{11}$ A/m$^2$. The collapse of a single skyrmion occurs at $J$ larger than $10.8 \times 10^{11}$ A/m$^2$ but when there is interaction with the trailing skyrmion-A, it exerts a force opposite to the edge repulsive force, the collapse occurs at lower $J$ starting from $8.8 \times 10^{11}$ A/m$^2$. This reduction of the annihilation current density is due to the interaction with skyrmion-A. By carefully investigating the skyrmion dynamics in a very short time scale, we noticed that the skyrmion-B burst from the core. For $J = 11 \times 10^{11}$ A/m$^2$, the core-collapse of skyrmion B started at time $t_i = 1.2$ ns followed by a reduction of skyrmion size in just a duration of 29 ps (Fig. 6.a-c). After 27 ps, the burst of the skyrmion started from the core with a counter-clockwise chirality like the collapse of the star supernova (Fig. 6.b). Interestingly, both the size and the position of skyrmion-A did not change and its interaction with skyrmion-B was favoring its survival by avoiding touching the edge of the device. Fig. 6(c) shows the fast expansion of the remaining from skyrmion-B, which occurred in just 2 ps. However, for $J = 13 \times 10^{11}$ A/m$^2$ we observed a core-collapse for the leading skyrmion-B [32–34] and a bubble-collapse for the trailing skyrmion-A (zone 2 in Fig.

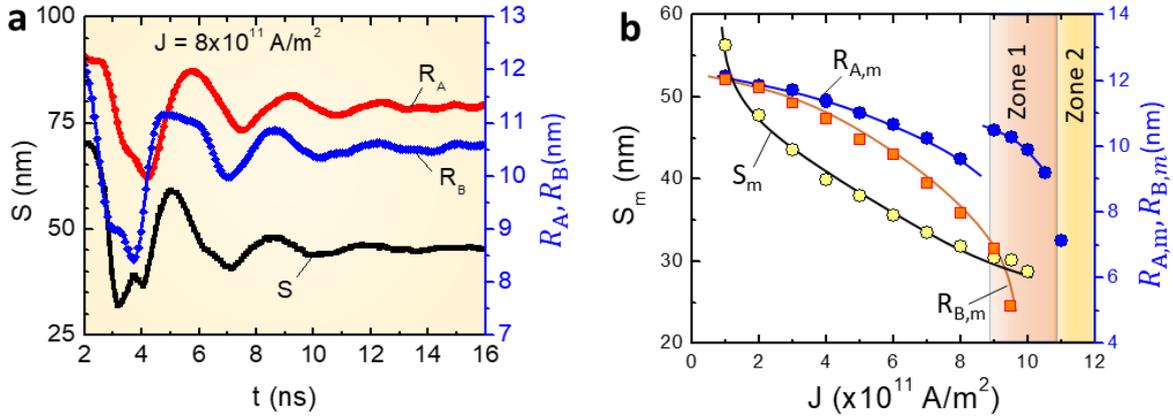

**Figure 4 | The separation and radius of interacting skyrmions.** (a) The skyrmion-skyrmion separation and their mutual radius versus time for $J = 8 \times 10^{11}$ A/m$^2$. (b) The minimum separation and the minimum radius of each skyrmion for different applied current densities. In zone 1, skyrmion-B is annihilated after a certain time when it is too close to the edges while in zone 2, the remaining skyrmion-A is annihilated (See supplementary Video S$_1$).

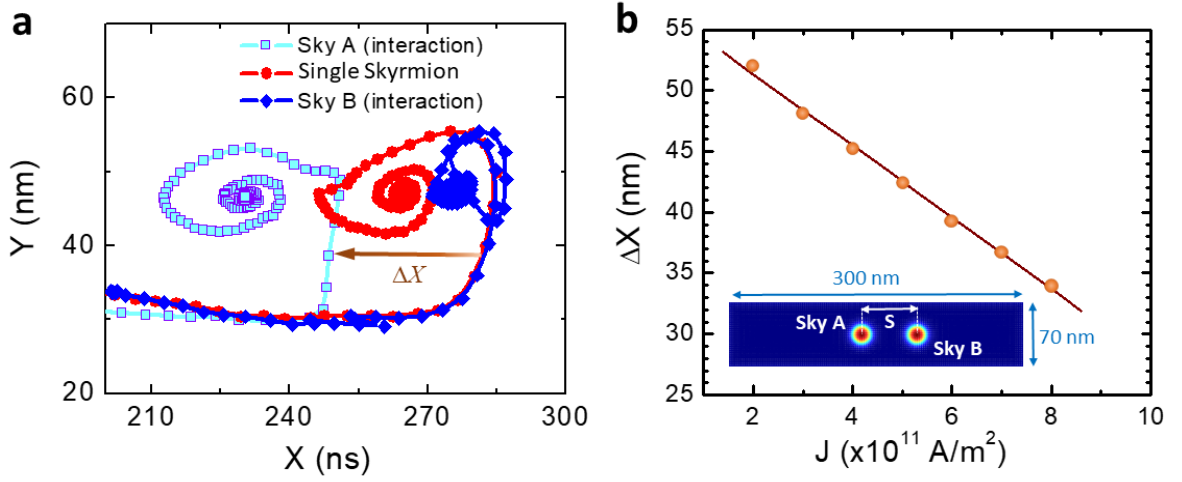

**Figure 5 | The separation and radius of interacting skyrmions.** (a) The motion of single skyrmion (middle path) and two interacting skyrmions at $J = 8 \times 10^{11}$ A/m$^2$. The shift $\Delta x$ indicates the effect of skyrmion-skyrmion interaction affecting the trailing skyrmion-A while the leading skyrmion-B is compressed due to the interaction with skyrmion-A and the two edges. (b) the plot of $\Delta x$ versus the current density showing a linear dependence. Inset is an image of the nanowire with the two skyrmions at the initial state.

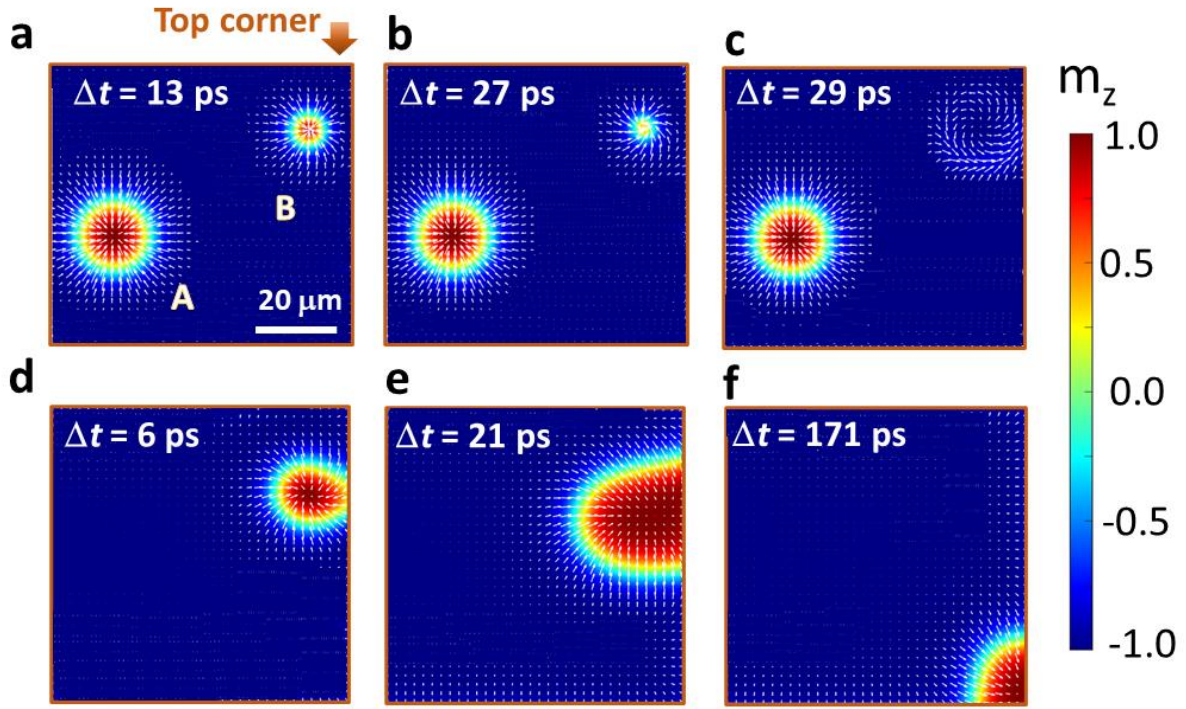

**Figure 6 | The Skyrmion annihilation mechanisms.** (a-b) Snapshots of the two interacting skyrmions at $J = 11 \times 10^{11}$ A/m$^2$.showing a core-collapse of skyrmion-B in less 30 ps time duration (d-e) Snapshots of the collapse of the trailing skyrmion-A after the collapse of Sky B at $J = 13 \times 10^{11}$ A/m$^2$. The bubble-collapse of skyrmion-A occurs at about 200 ps which is about 10 times slower than core-collapse in (a-c). The parameter $\Delta t$ is the time elapsed after the start of the annihilation.

4b). Fig. 6(d-f) is a series of images taken at different times for $J = 13 \times 10^{11}$ A/m$^2$ with the beginning of the annihilation which started at $t_i = 2.15$ ns when there is a slight touch at the right edge (Fig. 6d). From this state, one can see a progressive deformation of the skyrmion with an expansion like a bubble followed by a shrinkage toward the edge until complete collapse. It is worth mentioning here that the images shown in Fig. 6(d-f) are taken after the collapse of skyrmion-B as described in the case of $J = 11 \times 10^{11}$ A/m$^2$. The main reason for the collapse of skyrmion-A is there is no interaction with skyrmion-B and the driving force from STT is stronger.

**Conclusion**

In spintronic devices with more than one skyrmion, the interaction skyrmion-skyrmion has an impact of their dynamics. In this paper, we investigated the effect of the interaction between skyrmions within a nano-track on their respective paths and sizes and compared the results to the case of a single skyrmion (no-interaction). For a single skyrmion under STT, it exhibited a helical path which was enlarged with the current density *J*. Moreover, skyrmion radius reached its minimum when it is closer to the edge of the nano-track, especially, the corner. This is due to the net repulsive force from the right and top edges. For the case of the two interacting skyrmions, the radius of each one was very sensitive to their separation distance as a clear indication of their mutual repulsive force. The leading skyrmion which was driven by STT to the right edge and repealed from the trailing one shrunk more than in a non-interacting case. More interestingly, the annihilation of the leading skyrmion occurred at *J* lower than required for annihilating the trailing one. Because of the existence of difference forces acting differently on each skyrmion, their annihilation mechanisms were different.

**Model and simulation**

Numerical solutions for the skyrmions dynamics were evaluated using micromagnetic simulation namely Mumax3 [35] which resolve the magnetization at the lattice site and time evolution of the magnetization by the Landau-Lifshitz-Gilbert (LLG) equation:

$$\frac{d\vec{m}_i}{dt} = -|\gamma|\left(\vec{m}_i \times \vec{H}_i^e\right) + \alpha\left(\vec{m}_i \times \frac{d\vec{m}_i}{dt}\right) + \frac{Pa^3}{2eM_s}(\vec{j}(\vec{r}).\vec{\nabla})\vec{m}_i$$
$$- \frac{Pa^3\beta_J}{2eM_s}[\vec{m}_i \times (\vec{j}(\vec{r}).\vec{\nabla})\vec{m}_i] \qquad (1)$$

where $\vec{m}_i$, $\gamma$ and $\alpha$ are the normalized magnetization unit vector, the gyromagnetic ratio and Gilbert damping constant originating from spin relaxation, respectively. The effective field $\vec{H}_i^e$ acting on the local magnetization includes the magnetostatic energy, total exchange energy, DMI energy and anisotropy energy terms. The first and second terms in the right of Eq. 1 are the precession and damping terms while the third and fourth terms describe the coupling between the spin and the spin-polarized electric current $\vec{j}(\vec{r})$ via spin-transfer torque (STT) and via non-adiabatic effects respectively. The parameters *P*, *a* and *β*$_J$ are the spin polarization of material, lattice constant and the strength of the non-adiabatic torque, respectively.

The intrinsic material parameters of the layer used in this study are saturation magnetization $M_s$ = 500 kA/m, exchange stiffness constant $A_{ex}$ = 15 pJ/m, interfacial Dzyaloshinskii-Moriya strength *D* = 3.3 mJ/m² , with perpendicular magnetocrystalline anisotropy energy $K_u$ = 0.8 MJ/m³, Landau-Lifshitz damping constant α = 0.1. The nano-track has a thickness of 2 nm, a length *L* of 300 nm and a width *W* of 70 nm. The calculated magnetostatic exchange length $l_{ex} = \sqrt{2A/\mu_o M_S^2}$ and magnetocrystalline exchange length $\delta = \sqrt{A/K_u}$ in this system are 9.77 nm and 4.33 nm, respectively. The investigated device was discretized into equal cubes of 2 nm length and the grid size was set to 256 × 128 × 1, which is smaller than the exchange length to ensure a reasonable numerical accuracy in the relaxation process and time-dependent magnetization dynamics. The calculation was conducted in the absence of thermal fluctuation and performed without applying an external magnetic field.

It has been reported that the skyrmion is stable for $W > 2R$ otherwise, it will immediately collapse [36]. In the second part of this study, two skyrmions, with the same topological charge were initially positioned near each allowing the system to relax, then a spin-polarized current with a polarization of 0.4 is applied from right to left. The non-adiabaticity factor $B_J$ of the STT was fixed to a value of 0.2 and the net-pulsed current was applied to move the skyrmion in the opposite direction. The skyrmion dynamics were recorded and studied for the cases of a single skyrmion and two interacting skyrmions.

## Acknowledgments

The authors would like to acknowledge the support from HMTF Strategic Research of Oman (grant no. SR/SCI/PHYS/20/01).